\documentclass[apj]{emulateapj}
\usepackage{apjfonts,amsmath}
\usepackage{graphicx}

\shorttitle{Individual Time-Steps in SPH Simulations}
\shortauthors{Saitoh and Makino}

\begin{document}

\title{A Necessary Condition for Individual Time-Steps in SPH Simulations}

\author{Takayuki \textsc{R.Saitoh}\altaffilmark{1},
        Junichiro \textsc{Makino}\altaffilmark{1,2}
}
\altaffiltext{1}{Center for Computational Astrophysics,
National Astronomical Observatory of Japan, Mitaka, Tokyo 181-8588, Japan}
\altaffiltext{2}{Division of Theoretical Astronomy,
        National Astronomical Observatory of Japan, 2--21--1 Osawa,
        Mitaka-shi, Tokyo 181--8588; 
        and School of Physical Sciences, Graduate University of Advanced Study (SOKENDAI)}
\email{saitoh.takayuki@nao.ac.jp,saitoh.takayuki@cfca.jp}

\received{2008 Aug 5}
\accepted{2009 Apr 9}

\begin{abstract}
We show that the smoothed particle hydrodynamics (SPH) method, used with
individual time-steps in the way described in the literature, cannot handle
strong explosion problems correctly.  In the individual time-step scheme,
particles determine their time-steps essentially from a local Courant
condition.  Thus they cannot respond to a strong shock, if the pre-shock timescale
is too long compared to the shock timescale.  This problem is not severe in
SPH simulations of galaxy formation with a temperature cutoff in the cooling
function at $10^4~{\rm K}$, while it is very dangerous for simulations in
which the multiphase nature of the interstellar medium under $10^4~{\rm K}$
is taken into account.  A solution for this problem is to introduce a
time-step limiter which reduces the time-step of a particle if it is too
long compared to the time-steps of its neighbor particles.  Thus this kind
of time-step constraint is essential for the correct treatment of explosions
in high-resolution SPH simulations with individual time-steps.
\end{abstract}

\section{Introduction}

Hierarchical (individual) time-step method \citep[e.g.,][]{McMillan1986,
HernquistKatz1989, Makino1991IndividualTimeStep} is widely used in
simulations of galaxy formation and star formation based on smoothed
particle hydrodynamics (SPH) method \citep{Lucy1977, GingoldMonaghan1977}.
This method allows particles to have different time-steps, and can
significantly reduce the total calculation cost when there is a large
variation in the timescales of particles. Almost all implementations of the
individual time-steps method used for particle systems violates Newton's
third law (see \citeauthor{FarrBerschinger2007}
\citeyear{FarrBerschinger2007} for one of exception).  As long as physical
quantities are integrated with sufficient accuracy, this violation is not a
severe problem.

However, it is not always possible to maintain the accuracy.  To our
knowledge, all existing implementations of individual time-step for SPH rely
on the determination of the time-step at the end of previous time-steps.
Therefore, if something unforeseen occurs during the time-step of one
particle, the particle might fail to catch that event, resulting in a large
integration error.  A supernova (SN) explosion is an example of such an
event.  A SN generates a small amount of very hot gas ($T \sim 10^8~{\rm
K}$) in a large clump of cold gas ($T \sim 10~{\rm K}$).  The difference in
the time-steps of the hot gas and surrounding cold gas particles becomes
quite large (typically the difference reaches $\sim 10^3$).  Thus, hot gas
particles step forward $\sim 10^3$ or more time-steps before neighboring
cold gas particles respond to the SN event.  This means that the evolution
of both hot and cold gas particles is completely wrong, since the
surrounding cold gas particles do not react the explosion for a duration
much longer than the timescale in which the blast wave would propagate the
inter-particle distance.  This problem is not severe for ordinary SPH
simulations of galaxy formation because of the temperature cutoff in cooling
functions at $10^4~{\rm K}$, while it becomes very serious for simulations
involving the multiphase nature of the interstellar medium under $10^4~{\rm
K}$, because the mach number can be very high.

This problem of sudden change occurs in any dynamical simulation with
individual time-steps, as long as the time-steps are determined with the
usual explicit method.  In principle, a fully implicit method in which the
time-step itself is also determined implicitly \citep{Makino+2006} or a
method which satisfies Newton's third law \citep{FarrBerschinger2007} can
solve this problem, but there are no implementations of such methods for SPH
simulations yet {\footnote {Recently, \cite{Springel2009} has developed a
mesh-based scheme employing individual time-steps where the smaller
time-step is adopted for integrations between neighboring meshes. This
scheme does satisfy Newton's third law.}}.

In simulation of star clusters, the SN and its kick introduces a sudden
change in the orbit (and mass) of the exploded star. Here, a rather simple
prescription in which either all stars or at least nearby stars are
synchronized to the time of explosion and restart the integration has been
used. This prescription, at least the version which synchronizes all
particles in the system, is impractical for $N$-body/SPH simulations of
galaxies, because the number of SN events and therefore the increase in the
calculation cost is too great.

We propose a simple limiter for hydrodynamical time-steps in the individual
time-step method which mitigates this problem.  With this limiter the
behavior of an explosion integrated by individual time-steps becomes
essentially the same as that integrated by global time-steps.  In \S 2, we
describe this limiter, and in \S 3 we report the result of numerical
experiments.

\section{The Time-step Limiter}

We denote the time-step of an $i$-th particle as $dt_i$ and that of a
neighbor particle, with index $j$, as $dt_j$.  The basic idea of our limiter
is to enforce the following conditions:
\begin{eqnarray}
dt_i \le f~dt_j &&  ~{\rm{and}},  \\
dt_j \le f~dt_i && , 
\end{eqnarray}
where $f$ is an adjustable parameter. We found $f = 4$ to give good results,
from the perspective of the total energy and linear momenta conservations,
without significant increase in the calculation cost (see Table
\ref{tab:conservation} in \S \ref{sec:results}).  Note that too large values
of $f$, say $f > 4$, lead to the violation of the Courant condition.  It is
essential that the time-step of a particle $j$ shrinks when the time-step of
its neighbor particle $i$ suddenly shrinks by a large factor.  Thus particle
$i$ should let particle $j$ respond to the change of its time-step.  The key
point of the limiter is to reduce effects of the violation of Newton's third
law.  Although our approach does not guarantee the third law, we can achieve
sufficient accuracy by controlling the violation via the control parameter,
$f$.

We implement this limiter in the following way: when integrated, particle
$i$ sends its new $dt_i$ to its neighbor ($j$-th) particles.  The particles
receiving the time-step of particle $i$ compare it with their time-steps,
$dt_j$.  When $dt_j > f~ dt_i$ they shorten their time-steps so that they
satisfy the relation $dt_j = f~dt_i$.

Note that this reduction of the time-step of particle $j$ is only possible
if the times of two particles $t_i$ and $t_j$ satisfy the condition
\begin{equation}
  t_i \ge t_j + f~dt_i.
\end{equation}
If the above condition is not satisfied, the reduction of time-step results
in the new time of particle $j$ before the current time of particle $i$,
requiring the backward integration of the entire system.

\begin{figure*}
\begin{center}
\epsscale{1.0}
\plotone{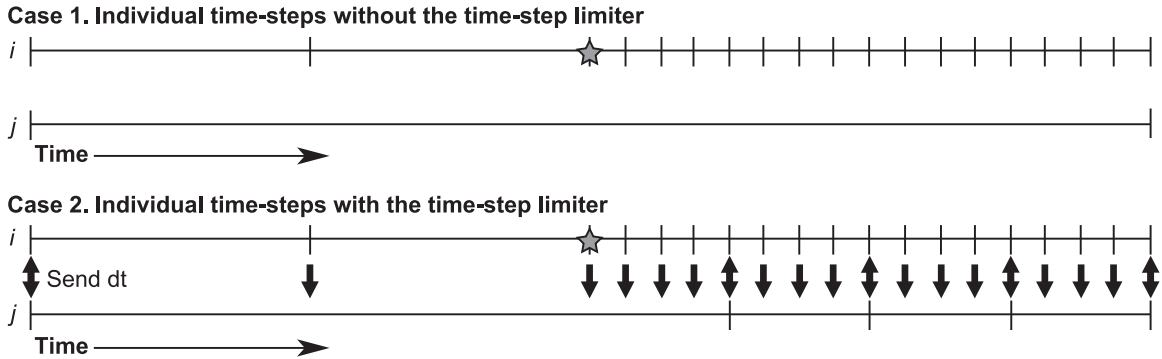}
\caption{Schematic pictures of individual time-step methods without and with
the time-step limiter (case 1. and 2.). Particle $i$ suddenly shrinks its
time-step ($dt_i$) at the time indicated by the star symbol.  In case 1.,
particle $j$ does not respond to the change of $dt_i$.  In case 2.,
particles $i$ and $j$ always send their time-steps ($dt_i$ and $dt_j$) to
their neighbors when they become active.  If the minimum time-step of
neighbors is shorter than its own time-step by a factor $f$ (see text), a
particle shortens its time-step.
\label{fig:Schematic}
}
\end{center}
\end{figure*}

In this case, the new time of particle $j$ is set to a value that is
consistent with the current system time ($t_i$), and the time-step is set to
the difference between particle' current time and new time.  In Figure
\ref{fig:Schematic}, we show schematic pictures of the traditional
individual time-steps method and our implementation of the individual
time-steps method.

\section{Numerical Experiments}
\subsection{Initial Setups and Method}

In order to verify the effect of the limiter, we performed two explosion
tests. The first test deals with hydrodynamics only while the second one
treats both hydrodynamics and self-gravity.  For each test, we performed
three runs: (a) integrated with global time-steps (hereafter Run a), (b)
traditional individual time-steps (case 1. in Fig. \ref{fig:Schematic}; Run
b), and (c) individual time-steps with the limiter (case 2. in Fig.
\ref{fig:Schematic}; Run c).

For the first test we adopted the Sedov problem, which is a pure
hydrodynamical evolution of the point-like explosion in the cold and
homogeneous ambient medium \citep{Sedov1959book}.  We prepared a glass
distribution with $64^3$ SPH particles.  We enhanced the thermal energy of
the central 32 particles in an SPH manner.  The total thermal energy of the
explosion is set to be unity.  The internal energy of the ambient gas
particle is set to be $10^{-6}$ of that for the hottest gas particle,
although the original Sedov problem adopted a zero-energy background.
Hence, the energy ratio is comparable with SN explosion, which generates a
small amount of very hot gas ($T \sim 10^7~{\rm K}$), in cold ($T \sim
10~{\rm K}$) ambient gas. In this configuration, the position of the shock
front at the time $T$ is $\sim 1.15 \times T^{2/5}$, when we assume the
adiabatic index $\gamma = 5/3$.

\begin{figure}
\begin{center}
\epsscale{0.9}
\plotone{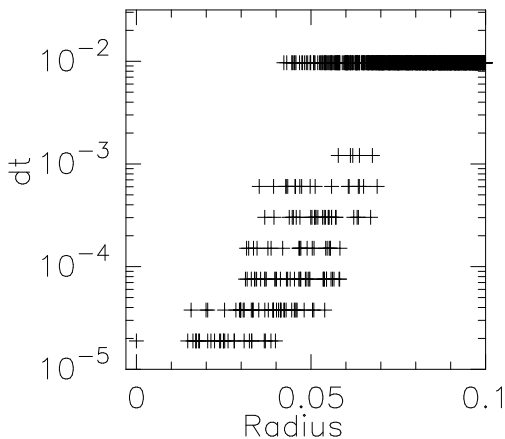}
\caption{The distribution of initial time-steps for the Sedov problem with
individual time-steps as a function of the radius. 
\label{fig:init_dt}
}
\end{center}
\end{figure}

In figure \ref{fig:init_dt}, we show the initial distribution of time-steps
for the Sedov problem with individual time-steps as a function of the
radius.  The difference in time-steps between the hottest particle and the
ambient particles is $\sim 10^3$.  Note that time-steps of particles in the
contact region surrounding the hot region are smaller than those of distant
ambient particles ($dt = 0.009$), since we estimate these time-steps after
the internal energy of particles are specified.  This distribution of
time-steps is quite moderate compared to that in practical simulations just
after SN explosions, in which only the particles which receive the energy
have small time-steps.

We then perform an explosion test of a cold cloud including the
self-gravity.  For this test, we used the particle distribution of the
three-dimensional collapse test \citep[e.g.,][]{Evrard1988} at $T = 3$ where
the system becomes the state of virial equilibrium with the total energy of
$E \sim -0.6$.  We added the thermal energy to the central 32 particles in
an SPH manner in order to drive the explosion of the system.  The total
energy of the system was set to be $E = 10$.

We designed this test to mimic the SN energy input in high-density
star-forming molecular cloud.  The total binding energy of a cloud with the
total mass of $10^4~M_{\sun}$ and the size of $10~{\rm pc}$ \cite[see Table
1 in][]{BerginTafalla2007} is around $-10^{48}~{\rm ergs}$, while the energy
input of a type II SN is $10^{51}~{\rm ergs}$.  Therefore, if we scale the
energy accordingly, the energy input should be $\sim 600$. We used a much
smaller value, so that the breakdown of the traditional method is less
severe.  We reset $T = 0$ at the initial state.  The number of SPH particles
used here is 30976.  The gravitational softening, $\epsilon$, is set to be
$0.05$.

We use the standard SPH method \citep[e.g.,][]{Monaghan1992}.  The
asymmetric form is adopted for the derivation of the internal energy.  The
signal velocity based artificial viscosity term \citep{Monaghan1997} is
used.  The viscous coefficient, $\alpha$, is set to be two for the first
test and unity for the second test.  We determine the interaction (kernel)
size of each SPH particle to keep the neighbor number to $32 \pm 2$.
Gravity is solved by the Tree with GRAPE method
\citep{Makino1991TreeWithGRAPE} with an opening angle of 0.5.

The time-integrator uses the leap-frog method.  In runs of pure
hydrodynamics, the time-step for individual particles is calculated as the
smaller of the ones calculated using the signal velocity based method
\citep{Monaghan1997,Springel2005} and the acceleration time-step; $0.3
\times (2 h / |a_{\rm h}|)^{0.5}$ \citep{Monaghan1992}. Here $h$ and $a_{\rm
h}$ are the kernel size and the acceleration caused by the pressure
gradient, respectively.  The coefficient for the Courant condition, which is
estimated by the signal velocity based method, is set to be 0.3.  When we
consider self-gravity, we modify the acceleration time-step taking into
account the gravitational acceleration, $a_{\rm g}$, and $\epsilon$; $0.3
\times (\min(2~h,\epsilon) / |a_{\rm h} + a_{\rm g}|)^{0.5}$.

\subsection{Results} \label{sec:results}

\begin{figure*}
\begin{center}
\epsscale{1.0}
\plotone{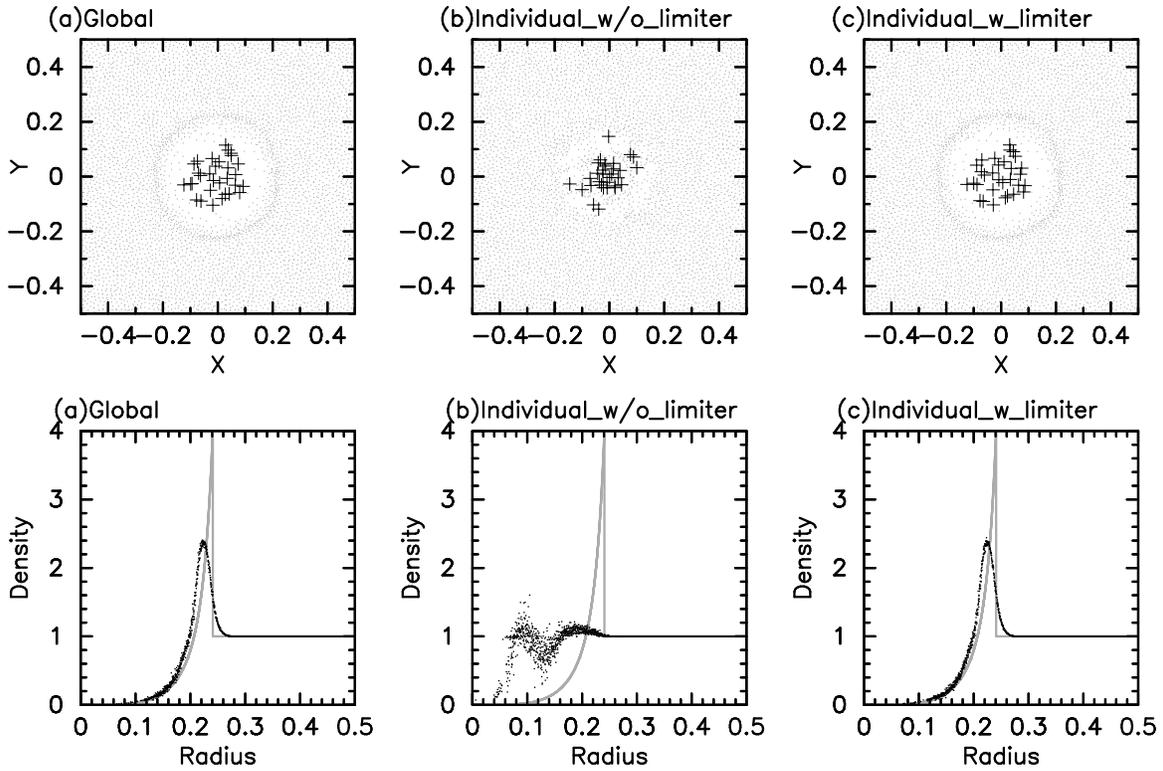}
\caption{Left, middle, and right column of panels show the evolution of
particles integrated with (a) global time-steps, (b) individual time-steps,
and (c) individual time-steps and the time-step limiter, respectively.  The
upper row shows particle distributions for three runs at $T = 0.02$.  Points
indicate projected positions of SPH particles in sliced regions ($|z| <
0.01$) while crosses show those of initially heated SPH particles throughout
whole regions.  The lower row shows radial density profiles for three runs
at $T = 0.02$.  Points show geometrical means of densities for every radial
bin with the width of $0.005$ from the origin of coordinates.  Solid (gray)
curves are the analytic solution for the zero-energy background obtained by
\citet{Sedov1959book}.
\label{fig:sedov}
}
\end{center}
\end{figure*}

Figure \ref{fig:sedov} shows the result of the Sedov problem. In the top
row, particle distributions of sliced regions ($|z| < 0.01$) are shown, and
in the bottom row radial density profiles are shown. Run (a) (the left panels
in Fig. \ref{fig:sedov}) shows a clear spherical structure and its density
profile traces the analytic solution very well.

Run (b) (the middle panels), however, displays penetrations of the hot gas
particles into the ambient medium and fails to reproduce the analytic
solution.  This is because the time-steps of the gas particles composing the
ambient medium are very long ($dt \sim 10^{-2}$) compared with those of the
initially heated particles (the minimum value is $dt \sim 10^{-5}$) and
therefore the ambient medium has had only two time-steps before the time of
the snapshot ($T = 0.02$).  They could not respond correctly to the
expansion of the central hot region.

Run (c) (the right panels) does not show this problem.  In this run, the
penetration of the hot gas particle occurring with the traditional method is
suppressed perfectly and the analytic solution of the Sedov problem is
reproduced very well.  This is because the short time-steps propagate to
surrounding particles quickly enough.  As a result, cold gas particles can
respond to the pressure of the hot gas particles.  Consequently this run can
reproduce the analytic solution of the Sedov problem, like Run (a).  We can
conclude that this limiter or a similar criterion is necessary for the
simulations of SPH with individual time-steps involving strong explosions.

The total (kinetic + thermal) energy and linear momenta of the system at the
final phase ($T = 0.04$) are shown in Table \ref{tab:conservation}.  Run (a)
conserves the total energy to $\sim 10^{-3}$ and the linear momentum with a
very high precision.  While Run (b) shows violations of conservation of both
the total energy and the linear momentum because the time-steps of cold gas
particles are far too long to resolve the propagation of the blast wave.

Introduction of the time-step limiter greatly reduces these errors.  Runs
(c) conserve the total energy to a level similar to that of Run (a).  In
addition, the conservations of the linear momentum in these runs are
improved by four orders of magnitudes compared with that in run (b).  When
we compare runs with $f = 2$ and $4$, we found the run with $f = 2$ shows
results slightly better in conservations.  The distribution of particles and
radial profiles are indistinguishable.  Therefore we conclude that the
time-step limiter with $f=4$ is acceptable.

The calculation cost directly relates with the value of $f$; smaller $f$
leads to larger calculation cost.  Indeed, Run (c) with $f = 2$ takes
$1036~{\rm sec}$ for integration until $T=0.4$ whereas Run (c) with $f = 4$
requires only $489~{\rm sec}$.  Interestingly, calculation costs for Runs
(c) are smaller than that for Run (b).  This is because the hot bubble does
not expand correctly unless we introduce the limiter. 

\begin{table}
\begin{center}
\caption{Energy and momentum conservation and calculation cost.}\label{tab:conservation}
\begin{tabular}{lcccccc}
\hline
\hline
Run & $|(E_{\rm I}-E_{\rm F})/E_{\rm I}|$ & $|p_{\rm F}|_{\rm x}$ & $|p_{\rm F}|_{\rm y}$ & $|p_{\rm F}|_{\rm z}$ & Time (sec)\\
\hline
a      & 8.2e-4 & 1.4e-16 & 1.4e-15 & 7.7e-16 & 5646\\
b      & 7.2 & 5.4e-1 & 2.6e-1 & 3.1e-1 & 2068 \\
c($f=2$) & 7.6e-4 & 3.2e-5 & 3.1e-5 & 1.1e-5 & 1036 \\
c($f=4$) & 5.9e-3 & 1.7e-5 & 2.7e-5 & 2.2e-6 & 489 \\
\hline
\end{tabular}\\
\end{center}
Note. $E_{\rm I}$ and $E_{\rm F}$ are the total energy at $T=0$ (initial
phase) and $0.04$ (final phase), respectively.  $|p_{\rm F}|_{\rm x}$,
$|p_{\rm F}|_{\rm y}$, and $|p_{\rm F}|_{\rm z}$ indicate the absolute
values of linear momenta for $x$, $y$, and $z$ directions, respectively.
\end{table}

\begin{figure}
\begin{center}
\epsscale{1.0}
\plotone{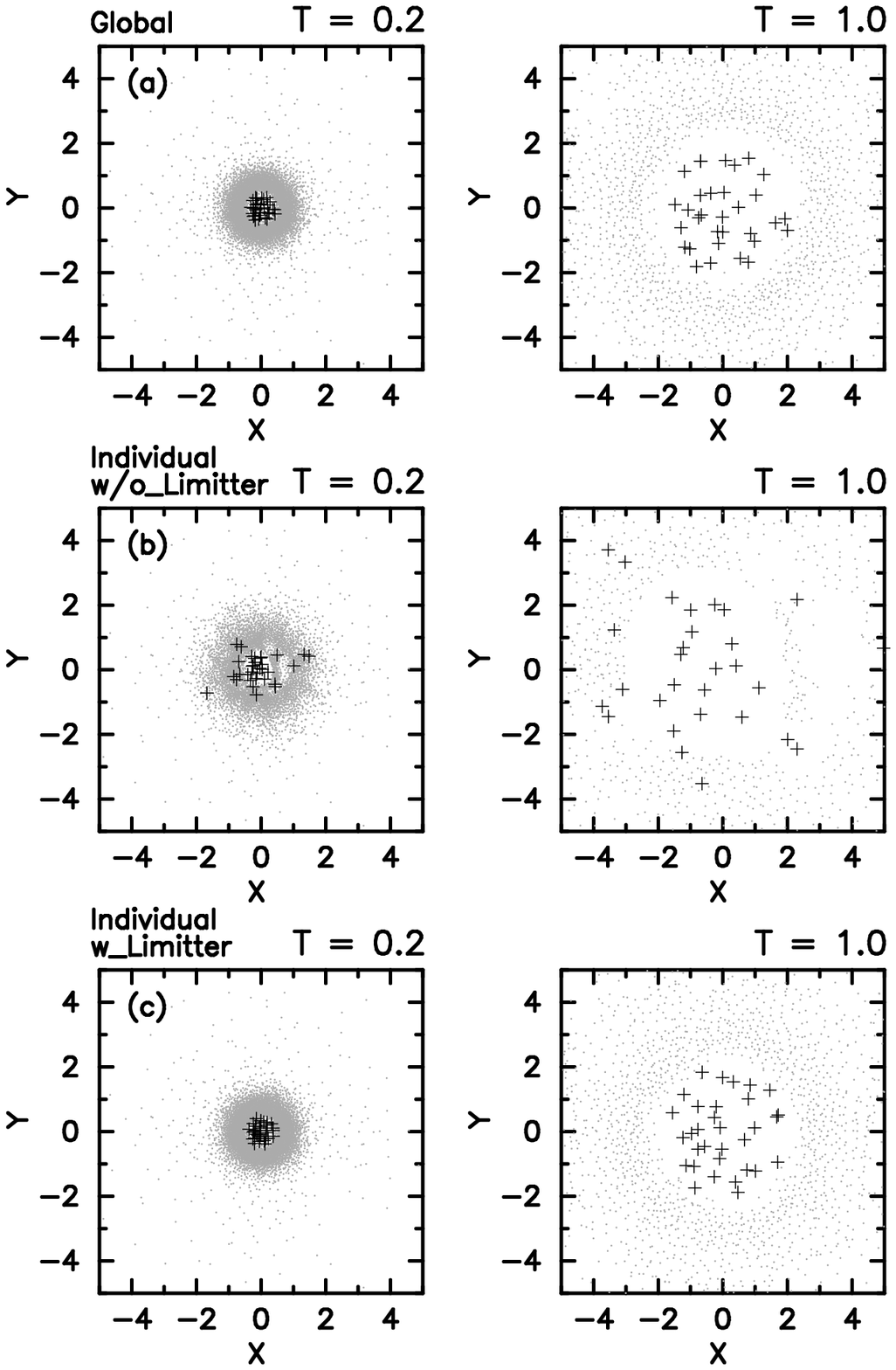}
\caption{Evolution of particles for three runs.  Upper, middle, and lower
rows of panels show the evolution of particles integrated with (a) global
time-steps, (b) individual time-steps, and (c) individual time-steps and the
time-step limiter, respectively.  Points and crosses indicate projected
positions of SPH particles in sliced regions ($|z| < 0.2$) and those of
initially heated SPH particles in whole regions, respectively.
\label{fig:slice}
}
\end{center}
\end{figure}

Figure \ref{fig:slice} shows sliced ($|z| < 0.2$) particle distributions for
Runs (a), (b), and (c) (top to bottom) at selected epochs, for the second model
with self-gravity.  Since the total (kinetic + thermal + potential) energy
of the system is positive and sufficiently large, the strong shock
propagates outward against the self-gravity resulting in an expanding shell.
Run (a) (top panels) shows this behaviors clearly.  The behavior of the
expansion for Run (b) (middle panels) surprisingly differs from that for run
(a).  Again, the penetrations of the initial hot gas particles into the cold
region are observed, and defects of the shell are shown in the middle-center
and right panels.  The introduction of the time-step limiter drastically
improves the results as can be seen in the bottom panels of Figure
\ref{fig:slice}.  There are no penetrations or defects of the spherical
shell-like shock which are visible in middle panels.

These results clearly indicate that the use of individual time-steps without
a time-step limiter is very dangerous for SPH simulations involving strong
explosion problems, such as high-resolution SPH simulations of galaxy
formation with a multiphase interstellar medium under $10^4~{\rm K}$.  On
the other hand, for traditional simulations of galaxy formation, the problem
might not be too severe since these simulations adopt the cut off
temperature of $10^4~{\rm K}$ for cooling function, and temperature
difference is at most around $100$. We strongly recommend that users of SPH who
investigate systems in which ``strong'' explosions take place with
individual time-steps implement some form of time-step limiter similar to
what we introduced here.

\acknowledgments

We thank the anonymous referee for his/her insightful comments and
suggestions, which helped us to greatly improve our manuscript.  We also
thank Takashi Okamoto for helpful discussion and  Kohji Yoshikawa for
providing us with a custom version of Phantom-GRAPE library. Numerical
computations were carried out on XT-4/GRAPE systems (project ID:g08a19) at
the Center for Computational Astrophysics, CfCA, of the National
Astronomical Observatory of Japan.


\end{document}